# AKN_Regie : une passerelle entre arts numériques et spectacle vivant

## *AKN_Regie : a bridge between digital and performing arts*

**Georges GAGNERÉ**

INREV-AIAC, Université Paris 8
georges.gagnere [at] univ-paris8.fr

**Résumé**. Parallèlement à la dissémination généralisée de l'informatique, on constate la persistance de frontières au sein des pratiques créatives, notamment entre les arts numériques et le spectacle vivant. Des traversées de ces frontières ont fait émerger la nécessité d'une appropriation commune des enjeux numériques. Fruits de cette appropriation, le dispositif AvatarStaging et sa dimension logicielle AKN_Regie seront décrits dans leur usage pour diriger des avatars sur une scène théâtrale mixte. Développé avec le langage visuel Blueprint au sein du moteur de jeu vidéo Unreal Engine de la société Epic Games, AKN_Regie offre un mode d'emploi accessible à des artistes non-programmeurs. Cette caractéristique servira à décrire deux perspectives d'appropriation de l'outil : la perspective Plugin pour ces utilisateurs et la perspective Blueprint pour les artistes programmeurs qui veulent faire évoluer l'outil. Ces deux perspectives sont alors complétées par une perspective C++ qui aligne AKN_Regie sur le langage avec lequel le moteur est lui-même programmé. Les circulations entre ces trois perspectives sont finalement étudiées en s'inspirant de travaux sur l'écologie de l'intelligence collective.

**Abstract.** In parallel with the dissemination of information technology, we note the persistence of frontiers within creative practices, in particular between the digital arts and the performing arts. Crossings of these frontiers brought to light the need for a common appropriation of digital issues. As a result of this appropriation, the AvatarStaging platform and its software dimension AKN_Regie will be described in their use to direct avatars on a mixed theatre stage. Developed with the Blueprint visual language within Epic Games' Unreal Engine, AKN_Regie offers a user interface accessible to non-programming artists. This feature will be used to describe two perspectives of appropriation of the tool: the Plugin perspective for these users and the Blueprint perspective for programming artists who want to improve the tool. These two perspectives are then completed by a C++ perspective that aligns AKN_Regie with the language with which the engine itself is programmed. The circulations between these three perspectives are finally studied by drawing on work on the ecology of collective intelligence.

## 1 Introduction

A l'heure où la révolution numérique semble avoir touché tous les secteurs de la société et tous les pays à travers la planète, son impact sur nos vies et nos pratiques socio-culturelles n'a probablement pas encore produit tous ses effets. Le changement de paradigme induit par l'émergence de l'informatique et de son déploiement électronique lié à la maîtrise de l'électricité dépasse probablement la mesure du précédant changement de paradigme qui a caractérisé l'émergence de l'alphabet, de l'écriture et des techniques de l'écrit autour du premier millénaire avant J.-C. (Ong, 1982). En complète immersion dans la culture de l'écrit, il nous est difficile d'imaginer des êtres humains aux capacités cognitives équivalentes mais sans recours à l'écrit pour communiquer et se développer. Cette culture de l'écrit qui a mis plusieurs siècles à émerger et dont l'acquisition nécessite plusieurs années est actuellement confrontée à de nouvelles pratiques reposant sur une manière de représenter le réel et l'acquisition de nouvelles capacités de communication utilisant du code informatique, mis en œuvre avec des ordinateurs.

La dissémination généralisée du code informatique et de ses extensions numériques prolifiques induit l'apparition de lignes de partage plus ou moins bien identifiables autour des enjeux d'appropriation du paradigme numérique au sens large, et dont la notion de frontières numériques constitue un outil d'investigation. Nous proposons de décrire une de ces lignes de partage dans le champ des pratiques créatives et plus précisément au niveau de la pratique du théâtre.

## 2 Frontières entre théâtre et numérique

Plusieurs décennies après la diffusion du paradigme numérique et la reconnaissance de son impact majeur sur les évolutions sociales, on constate l'expression d'inquiétudes profondes concernant « la domination de la technologie sur l'humain, les pouvoirs de transformation des réalités matérielles et corporelles en réalités virtuelles, la captation potentielle des qualités vibratoires et sensibles du vivant et ses manipulations [qui] sont de nature à dématérialiser l'art dramatique » (Triffaud, 2008) et qui conduisent à préconiser une pratique théâtrale renouvelée qui préserverait les futurs artistes de toute contamination robotique. Ces inquiétudes font place à une forme de résistance revendiquée par rapport aux transformations en cours de la part de praticiens professionnels et directeurs d'école de théâtre. Ainsi Stanilas Nordey pouvait avancer en 2015 qu'il n'avait « pas l'impression que la question principale des jeunes acteurs aujourd'hui, – des acteurs, je dis bien –, soit celle des nouvelles technologies. La question de l'acteur, souvent, touche à la présence, la voix nue, le rapport au public direct. Je pense qu'il faut faire attention lorsqu'on qualifie cette évolution d'inéluctable. Le théâtre est là depuis un bon moment, le cinéma est là depuis cent ans, il est en train de mourir, comme la télévision. Le théâtre a cette force de résistance incroyable qui est dû à la mise en présence simple d'un homme, ou deux hommes ou deux femmes, avec une voix nue, dans un même lieu, dans un même espace, et c'est quelque chose qui a résisté pendant assez longtemps, parce qu'il y en a eu d'autres, des nouvelles technologies, avant l'arrivée de celles du moment. » (Féral, 2018).

Cette intervention était faite dans le cadre d'une table ronde en conclusion du colloque organisé par Josette Féral à l'Université Paris 3 Sorbonne Nouvelle et intitulé *Le corps en scène : l'acteur face aux écrans*. La manifestation réunissait plus de 80 participants venant de plus de 20 pays autour d'un état des lieux de la relation entre l'acteur de théâtre et les nouvelles technologies. Elle concluait elle-même un cycle

de réflexions sur la notion d'effets de présence, démarré en 2007 au Canada par Josette Féral et Louise Poissant dans un groupe de recherche au titre homonyme, Effets de présence. Après de multiples journées d'étude et colloques sur cette question de la présence et de la confrontation de l'acteur et du spectateur à de nouvelles entités technologiques déstabilisantes, on note que l'ouvrage issu de ce colloque conclusif, publié en 2018, constitue un sorte de reformulation des questions déjà présentes dans *Les Ecrans sur la scène*, un ouvrage marquant de la fin du siècle dernier, dirigé par Béatrice Picon-Vallin, et prenant acte des premiers effets induits par le paradigme du numérique (Picon-Vallin, 1998). Vingt ans plus tard, les difficultés du théâtre à s'approprier les mutations en cours demeurent.

Parallèlement à ce constat de défiance face à des mutations profondément déstabilisantes, on note l'émergence de Journées d'Informatique Théâtrale portées par Rémi Ronfard et l'équipe de recherche qu'il dirige à l'INRIA. Rémi Ronfard et Julie Valéro font la remarque suivante concernant la relation du théâtre au paradigme numérique : « on constate que les expressions les plus fréquemment retenues sont : « technologies numériques », « médias », « intermédialité », « réalité augmentée », « environnement numérique ». L'informatique, qui sous-tend toutes les techniques évoquées dans ces publications et manifestations, semble ainsi être le grand absent de ces intitulés et n'y est donc que très rarement mentionné comme tel. » (Valéro et Ronfard, 2020) Il leur semble alors nécessaire de créer un espace de rencontres « autour des problématiques, des questions, des difficultés que soulève l'association, au sein d'un même projet artistique ou universitaire, des sciences informatiques et de la pratique théâtrale ». On constate cependant qu'après la participation de deux laboratoires de recherche en informatique aux premières journées ayant eu lieu en 2020, l'équipe de Rémi Ronfard à l'INRIA, et celle du Laboratoire d'Informatique de Grenoble par la présence de Véronique Aubergé, seule l'équipe de Rémi Ronfard était présente aux secondes Journées d'Informatique Théâtrale ayant eu lieu en 2022. Et malgré le nombre important de productions théâtrales recourant aux nouvelles technologies, le chercheur en informatique passionné de théâtre constate qu'il semble « pourtant encore peser sur ceux qui osent imposer l'informatique comme moyen d'expression, le rendre visible au sein même de la représentation, un anathème : au titre de leur création sera quasi systématiquement apposé un qualificatif supplémentaire à « théâtre » : « art numérique », « nouvelles technologies » ou encore « théâtre visuel » […]. L'ordinateur a-t-il vocation à dénaturer l'art dont il s'empare, en le décalant, en le déportant vers d'autres horizons que l'on aurait encore des difficultés à entrevoir ? »

## 3 Traversées des frontières

Ces remarques me semblent significatives d'une problématique sur laquelle je travaille depuis le début des années 2000 et qui concerne les modalités d'appropriation de nouveaux outils créatifs dans le domaine théâtral, formulée en 2004 dans une réflexion intitulée « Le temps réel du temps réel » (Gagneré, 2022) et qui soulignait la nécessité d'adapter au jeu des comédiens l'utilisation des outils numériques temps réel permettant de manipuler les images, les sons et la lumière, en relation avec la dynamique créative propre aux répétitions de théâtre. L'enjeu me semblait alors être de trouver des manières d'agencer en temps réel des dispositifs complexes modifiant les flux temps réel des matériaux scéniques numériques, et donc d'adapter harmonieusement les possibilités algorithmiques de l'informatique temps réel aux spécificités créatives du spectacle vivant.

Ces questionnements s'inscrivent dans la perspective défendue par Gilbert Simondon d'une adaptation permanente des processus culturels aux développements techniques des sociétés et des êtres humains qui les composent : « La culture est ce par quoi l'homme règle sa relation au monde et sa relation à lui-même ; or, si la culture n'incorporait pas la technologie, elle comporterait une zone obscure et ne pourrait apporter sa normativité régulatrice au couplage de l'homme et du monde. […] Ce n'est pas la réalité humaine, et en particulier ce qui de la réalité humaine peut être modifié, à savoir la culture [...] qui doit être incorporée aux techniques comme une matière sur laquelle le travail est possible ; c'est la culture, considérée comme totalité vécue, qui doit incorporer les ensembles techniques en connaissant leur nature, pour pouvoir régler la vie humaine d'après ces ensembles techniques. » (Simondon, 1958) Le travail d'appropriation de la culture confrontée à l'émergence continue de nouvelles technologies est clairement mis en avant.

Le démarrage d'un dialogue entre les pratiques des arts numériques et du spectacle vivant autour de la figure de l'acteur a permis de délimiter des frontières concrètes entre l'avatar et l'acteur physique et de bricoler des traversées d'un monde à l'autre. Un résultat de ces traversées a été de découvrir les domaines de connaissances et de compétences qui s'ouvraient aux artistes travaillant de part et d'autre de cette frontière entre physique et numérique (Gagneré et Plessiet, 2015). L'étape suivante a consisté à imaginer des expérimentations favorisant les circulations à travers cette frontière à la fois sur le plan des outils, des compétences et des projets créatifs (Gagneré et Plessiet, 2016). Conduites selon les modalités de la recherche-création, les expérimentations ont permis de formaliser plus précisément les conditions d'une relation harmonieuse entre arts numériques et théâtre. Les retours d'expériences ont notamment souligné la nécessité d'une compréhension par les artistes du langage informatique sous-tendant la construction des avatars, alter-ego des acteurs physiques.

## 4 Contexte de création d'AKN_Regie

En 2018, j'ai formalisé sous le nom d'AvatarStaging un dispositif qui permettait de rassembler les enjeux du dialogue amorcé depuis 2015 entre compétences numériques et théâtrales (cf. figure 1). Cette formalisation accompagnait la description d'un atelier de pratique de direction d'avatar dans le cadre de la conférence internationale Motion Computing (Gagneré et Plessiet, 2018) et de la mise en place d'un site internet pour en documenter les ressources (Didascalie.net, 2023), et notamment la brique centrale de programmation informatique AKN_Regie. Le préfixe AKN marquait sa filiation directe avec l'architecture informatique AKeNe décrite dans (Gagneré et Plessiet, 2023). Cet outil numérique de manipulation d'avatar est une réponse concrète aux questionnements initiaux de 2004, aux expérimentations conduites depuis 2015 et au compagnonnage artistique que j'ai noué avec Cédric Plessiet, artiste numérique, programmeur informatique, membre de l'équipe de recherche Image Numérique et Réalité Virtuelle du laboratoire Art Contemporain Art des Images de l'université Paris 8 et collaborateur de plusieurs chorégraphes et metteurs en scène dans des projets de spectacle vivant. C'est aussi le fruit d'une prise de conscience concernant les conditions d'écriture des interactions artistiques entre entités physiques et numériques. Comment comprendre ces dernières sans se confronter à leur essence informatique ?

J'ai donc décidé de franchir la frontière qui sépare l'utilisateur d'interfaces logicielles du programmeur qui agence ces interfaces (Gagneré, 2020a). Il y a bien entendu différents degrés d'appropriation de cette compétence informatique et j'ai profité d'une caractéristique offerte par le moteur de jeu vidéo Unreal Engine, développé par Epic Games (2023a), pour basculer dans le monde de la programmation informatique. En 2015, Epic Games change son modèle économique et décide de mettre à disposition son moteur de jeu et ses sources C++ moyennant le versement de royalties sur les bénéfices réalisés par les jeux et les films qui l'utilisent. En pratique, cela revient à en rendre l'usage accessible sans frais dans les contextes de la pédagogie, de la recherche-création et des projets de spectacle vivant. Par ailleurs, Unreal Engine offre un moyen de faire de la programmation visuelle par l'intermédiaire du langage graphique Blueprint (Epic Games, 2023b), qui donne accès aux principales fonctionnalités du langage scripté C++, langage avec lequel le moteur est lui-même programmé. Les éléments de code programmés en Blueprint sont appelés des blueprints et présentent une interface de programmation nodale comparable à celle de l'environnement Max-MSP (Cycling, 2023), environnement avec lequel les premiers moteurs vidéo que j'ai utilisés au début des années 2000 étaient programmés. A l'époque je dialoguais avec un artiste programmeur ou un régisseur pour les intégrer dans mes mises en scène (Gagneré, 2015). Je suis désormais directement programmeur de l'outil que j'utilise pour diriger des avatars dans mes dernières mises en scène.

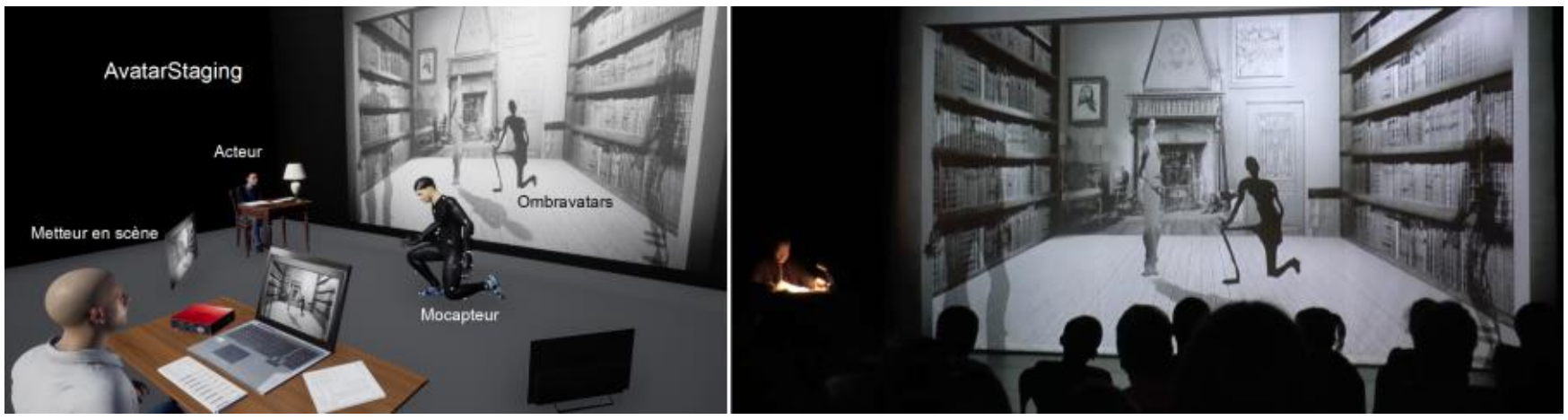

**Figure 1** *: A gauche : dispositif AvatarStaging. A droite : Extrait de* L'Ombre.

La figure 1 montre à gauche une vue 3D du dispositif AvatarStaging à l'occasion d'une répétition du spectacle *L'Ombre* (Gagneré, 2020b) dont on voit le résultat en situation de spectacle sur la photo de droite. A gauche, l'acteur à genoux portant une combinaison de capture de mouvement (mocapteur) contrôle un avatar sous forme de silhouette plate (ombravatar). L'animation est enregistrée pour être déclenchée ultérieurement pendant le spectacle par le conteur (à gauche de l'écran). Le dispositif a été conçu pour être manipulé par des utilisateurs non-programmeurs en informatique afin de respecter le cahier des charges des expérimentations conduites depuis 2015. Parmi les quelques utilisateurs de l'outil, Anastasiia Ternova, doctorante de l'équipe INREV-AIAC de l'université Paris 8 depuis 2019, l'a utilisé sur plusieurs recherche-créations et en a transmis le mode d'emploi à plusieurs équipes professionnelles (Ternova et Gagneré, 2023). Dans ces utilisations d'AKN_Regie, Anastasiia Ternova manipule l'interface et des éléments de l'environnement du moteur de jeu vidéo sans recourir à la programmation Blueprint. Elle se situe dans une zone d'appropriation de l'outil qu'il est intéressant de formaliser pour comprendre le contexte global d'utilisation d'AKN_Regie.

## 5 Perspectives sur l'utilisation d'AKN_Regie

En pratique, AKN_Regie offre un mode d'emploi de fonctionnalités qui permettent de contrôler des actions élémentaires de positionnement, de translation et de rotation sur des avatars enregistrés ou contrôlés en temps réel par un mocapteur, et cela d'une manière intuitive évitant à l'utilisateur de calculer par lui-même les opérations mathématiques qui rendent possibles des mouvements évidents à réaliser pour un être humain, comme se mettre sur une position donnée, tourner sur soi-même ou encore se translater vers l'avant. Ces opérations de manipulation se superposent aux mouvements propres des avatars et sont indispensables pour coordonner les actions scéniques dans l'espace 3D en relation avec la scène physique (Gagneré et Plessiet, 2018). Elles sont accessibles sous la forme de briques de programmation encapsulées (encore appelées nodes) comme SetAvatar, Cue et les éléments de la fenêtre de paramétrage Cuesheet, Devices ou encore AvatarsProps, tels que schématisés dans la figure 2.

Dans une démarche similaire, on peut manipuler ce qu'on appelle des « props », c'est-à-dire des accessoires numériques qui sont indépendants dans l'espace 3D ou bien ajoutés sur le corps d'un avatar. Ces props peuvent être tous les objets ou effets numériques programmables dans le moteur de jeu vidéo. On peut citer les maillages 3D, les lumières, les sons et les effets spéciaux avec particules. Leurs paramètres de contrôle sont circonscrits et rendus accessibles avec des nodes intitulés Cue, Set Prop et la fenêtre de paramétrage Avatars-Props selon le protocole de conduite d'effets scéniques d'AKN_Regie. La figure 2 représente ce que j'appelle une perspective Plugin sur l'outil et le positionne comme une fenêtre de programmation simplifiée au sein de l'environnement global Unreal Engine, dont la complexité sous-jacente est masquée. Un utilisateur non-programmeur doit seulement savoir installer le moteur de jeu, puis le plugin et suivre le mode d'emploi des fonctionnalités de contrôles des avatars et des props.

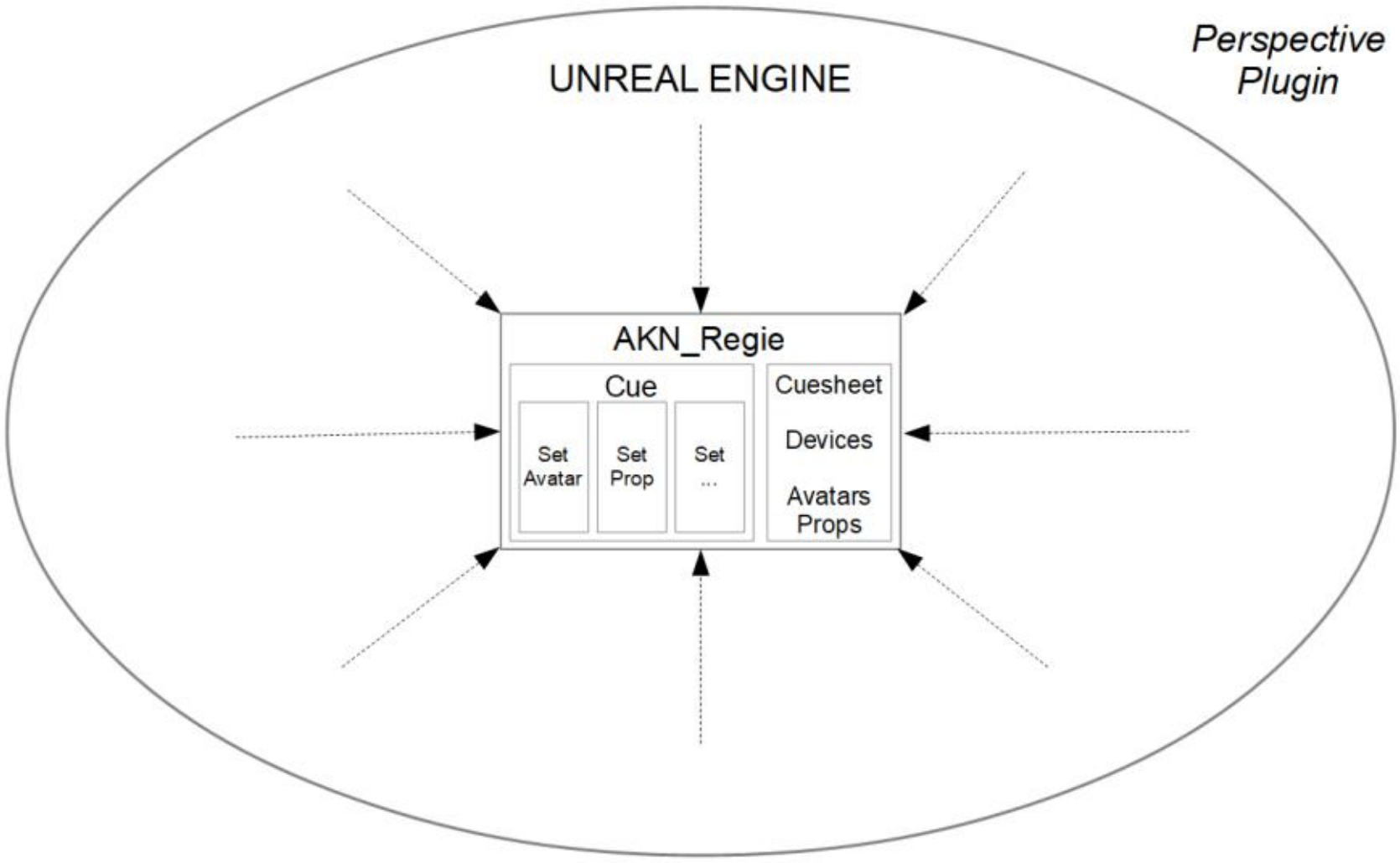

**Figure 2** : *Perspective Plugin sur l'utilisation d'AKN_Regie*

Il existe par ailleurs une documentation de la programmation blueprint d'AKN_Regie qui explique comment sont programmés les nodes de l'interface utilisateur à partir du langage Blueprint du moteur et de blueprints issus d'autres plugins ajoutés selon les besoins de programmation (Gagneré, 2023). Ces plugins ne sont pas nécessaires au fonctionnement standard du moteur. Il s'agit d'éléments de programmation complémentaires donnant accès à de nouvelles fonctionnalités, comme la réception d'informations provenant de logiciels tiers pour la capture de mouvement (plugin LiveLink) ou encore la prise en compte des signaux MIDI (acronyme du protocole de communication audio Musical Instrument Digital

Interface) des contrôleurs connectés à l'ordinateur sur lequel le logiciel opère (plugin MIDI). La figure 3 positionne ainsi AKN_Regie comme un ensemble de blueprints faisant appel à ces blueprints complémentaires et à d'autres blueprints du moteur comme ceux fabriqués avec les interfaces Editor Animation pour les mouvements ou Editor Niagara pour les effets visuels. L'architecture blueprint d'AKN_Regie est ainsi explicitée et potentiellement modifiable en utilisant le langage de programmation Blueprint. Il ne s'agit plus d'une boîte noire apparaissant à un utilisateur sous forme de plugin paramétrable et non modifiable.

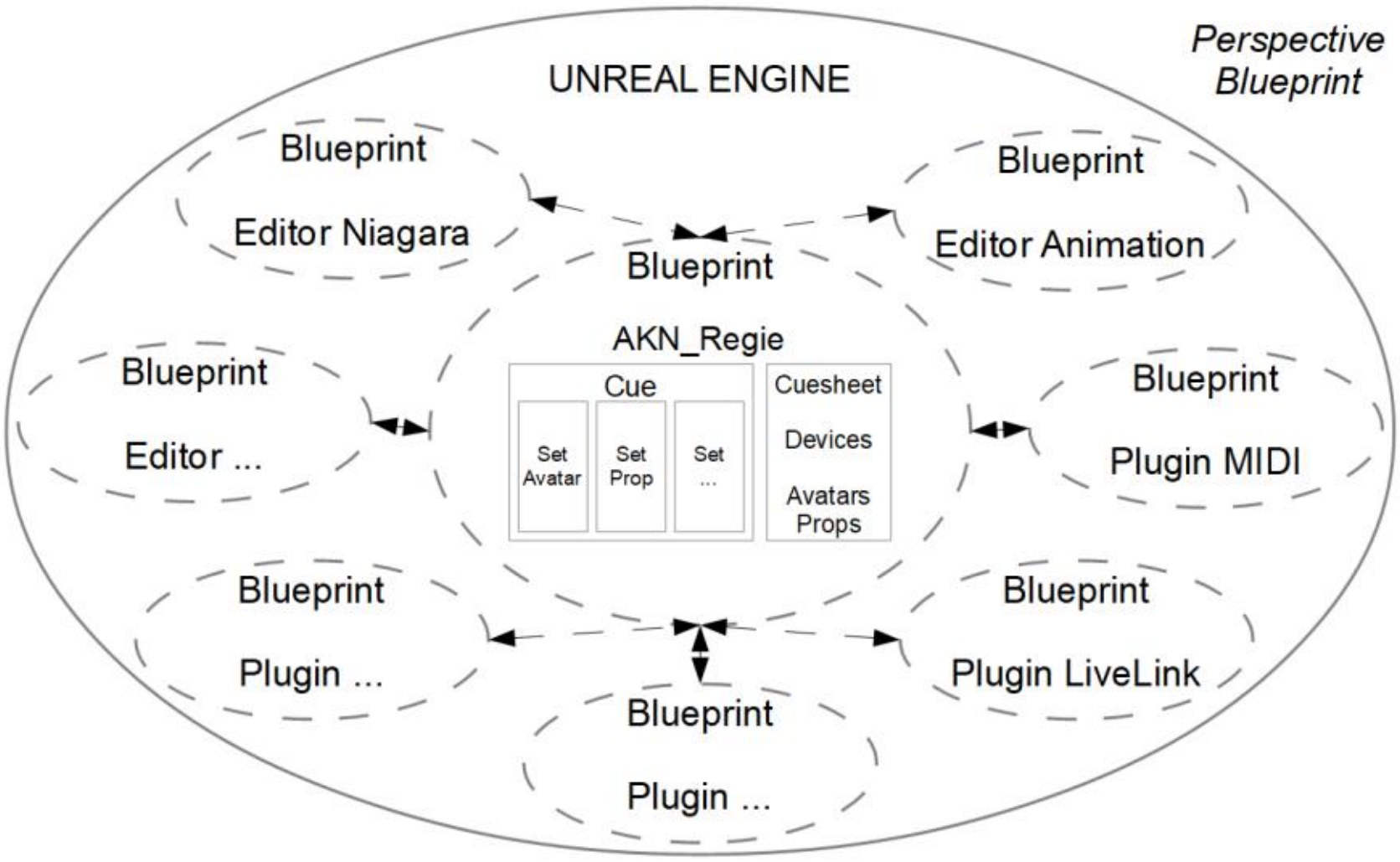

**Figure 3**. *Perspective Blueprint sur le développement d'AKN_Regie*

Nous formulons alors deux hypothèses à partir des usages réussis d'AKN_Regie en production, qui ont fait émerger le besoin de nouvelles fonctionnalités pour explorer l'évolution scénique des avatars. Premièrement, nous supposons qu'une connaissance de la manière dont AKN_Regie est programmé améliorerait la compréhension de la nature véritable des avatars, des potentialités d'agencement des fonctionnalités existantes, ce qui faciliterait la création de nouvelles fonctionnalités ou encore une amélioration de l'architecture globale de l'outil. L'approche simplifiée de la figure 2 utilisée pour agencer les fonctionnalités « utilisateur » de l'outil devrait alors être dédoublée par une compréhension de la programmation de ces mêmes fonctionnalités telles que schématisées par la figure 3. La figure 4 illustre à gauche la simplicité de l'interface plugin d'AKN_Regie. Le cadre A correspond aux paramètres Cuesheet/Devices/AvatarProps et le cadre B délimite 3 cues. La partie droite de la figure illustre la complexité de l'architecture des blueprints qui permettent de réaliser les actions sur les avatars et les props, et qu'il faudrait donc s'approprier pour être capable de naviguer selon la perspective Blueprint pour la faire évoluer.

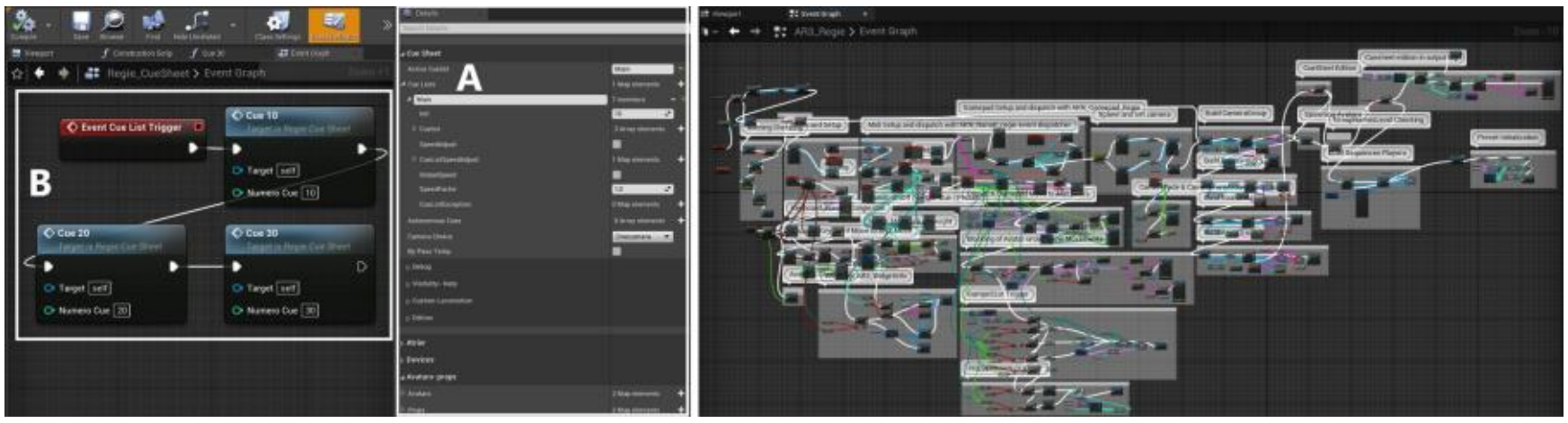

**Figure 4** *: A gauche : Interface donnant accès au plugin. A droite : Architecture des blueprints*

Une deuxième hypothèse consiste à approfondir la prise en compte de la complexité de programmation et propose que la circulation entre l'approche type conduite théâtrale de la perspective Plugin et l'architecture de la perspective Blueprint d'AKN_Regie soit étendue au langage même du moteur, le code C++. Cela consisterait à aborder à ce bas niveau de programmation les blueprints d'AKN_Regie, ceux des plugins complémentaires tels que LiveLink ou MIDI, ceux produits avec les interfaces Editor Anima ou Niagara et toutes les autres fonctionnalités du moteur. Parallèlement au fait de permettre une compréhension de la nature première des matériaux numériques utilisés de manière créative avec des êtres humains, la perspective C++ (figure 5) pourrait faciliter par exemple une approche de programmation par Design Patterns (Gamma *et al.*, 1994) et l'utilisation du Langage de Modélisation Unifié UML (Jakobson *et al.*, 1999). Elle permettrait d'optimiser la programmation visuelle en blueprints et d'intégrer des briques de programmation C++ extérieures au code existant d'Unreal Engine.

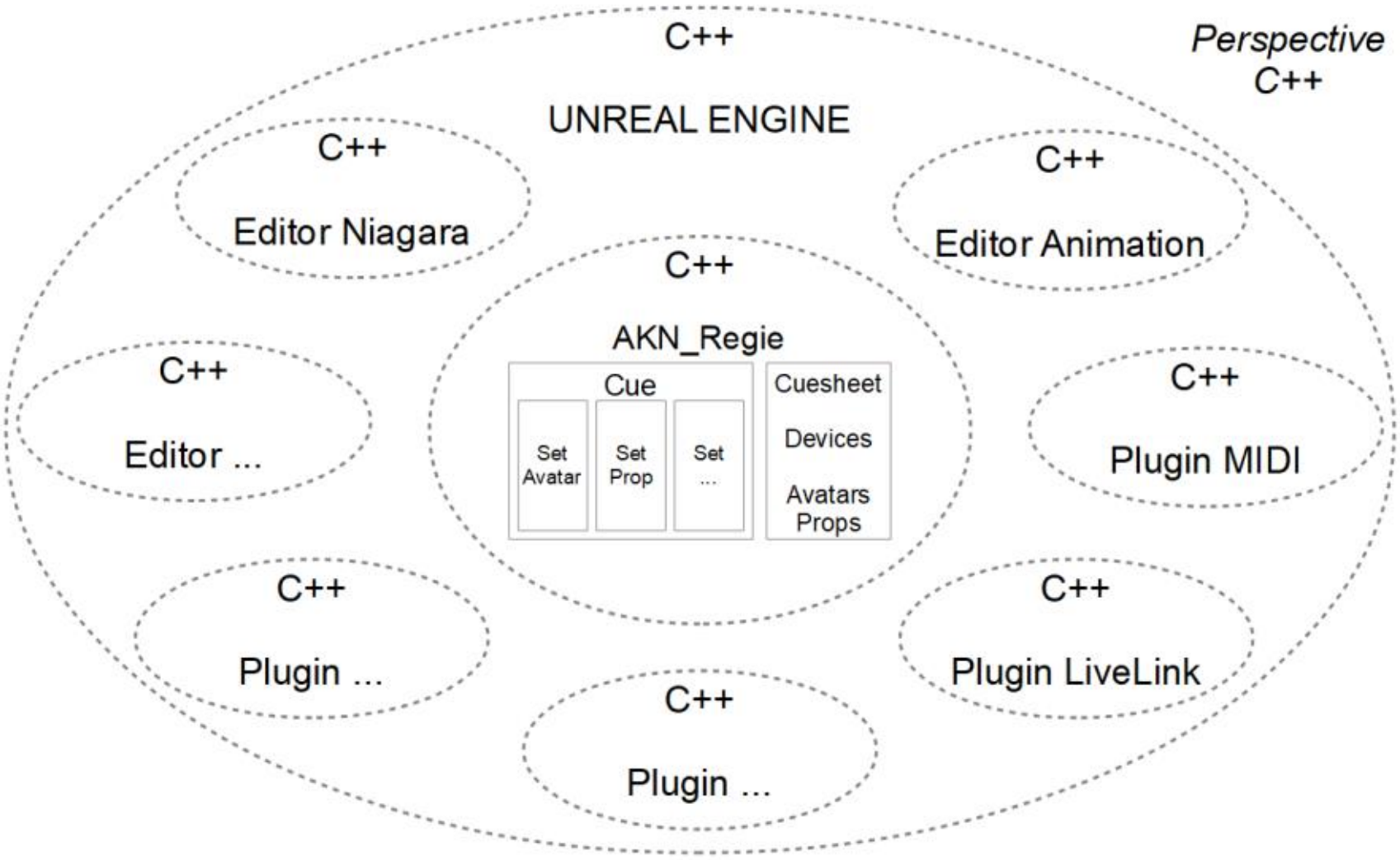

**Figure 5**. *Perspective C++ sur le développement d'AKN_Regie*

La circulation des points de vue entre les trois perspectives d'AKN_Regie demande à chaque fois de recontextualiser des connaissances dans un environnement différent. Quelle approche privilégier pour stimuler la pratique créative ? On peut supposer a priori que la perspective Plugin facilite un dialogue avec le metteur en scène et les acteurs, mais au détriment de la compréhension de la logique de fonctionnement des avatars. On peut aussi supposer que la perspective C++ permet d'inventer de nouvelles fonctionnalités sans garantir leurs utilisations en pratique sur une scène théâtrale avec des acteurs physiques. La question se résume à chercher une manière de faire cohabiter ces trois vues différentes sur l'utilisation des mêmes matériaux numériques à destination de la création artistique.

## 6 Conditions pour établir une passerelle

Le défi à relever pourrait s'inspirer des recherches de Samuel Szoniecky (2012) concernant l'élaboration d'une approche allégorique complétant les langages symboliques pour rendre compte de la complexité croissante des écosystèmes d'information. L'auteur décrit notamment trois approches complémentaires fournissant des outils pour développer l'intelligence collective, considérée par l'un de ses promoteurs comme « la capacité des groupes humains à collaborer sur le plan intellectuel pour créer, innover et inventer. Cette capacité peut être appliquée à n'importe quelle échelle, des petits groupes de travail jusqu'à l'espèce humaine en passant par des réseaux de toutes tailles. » (Levy, 2010). Il aborde ainsi la question de l'ingénierie des connaissances et du développement des ontologies proposée par Bruno Bachimont, la méthode de Conceptualisation Relativisée proposée par Miora Mugur-Schächter et enfin le programme de recherche de Pierre Lévy sur un métalangage pour l'économie de l'information (IEML pour Information Economy Metalanguage). Ces approches sont basées sur des conceptions symboliques du rapport au réel, dont le processus de codification a tendance par nature à figer le dynamisme de l'interprétation et de la signification, en tant que circulation de la forme au sens. Szoniecky propose alors de recourir à la notion d'allégorie et de s'inspirer des apports anthropologiques de Philippe Descola (2005) sur la notion d'ontologie analogiste qui combine l'association d'une pluralité des représentations de la matérialité physique et de l'intériorité spirituelle pour décrire les rapports entre humain et non-humain. « En effet, le contrôle strict d'une logique symbolique qui fonctionne dans une relation univoque entre la forme et le concept, a tendance à orienter nos sociétés numériques vers la performativité d'une société d'insectes. Pour amener vers d'autres voies nos sociétés numériques, il faut peut-être les concevoir dans la perspective d'écosystèmes d'information où la relation entre forme et concept puisse être multipliée par une infinité de points de vue. » (Szoniecky, *ibid.*, p. 198). L'auteur donne alors l'exemple d'un « agent allégorique » construit à partir de la notion de jardin et permettant de matérialiser la complexité du rapport à la connaissance.

Nous ne chercherons pas ici à déployer cet agent pour aborder la problématique de circulation entre les trois perspectives concernant l'outil AKN_Regie. Mais il nous semble que ce recours à la notion d'allégorie est une piste à explorer en termes d'intelligence collective pour enrichir le développement de l'outil. Il semble que la place du langage Blueprint et de sa représentation graphique nodale joue un rôle important d'appropriation des enjeux informatiques sous-jacents liés au langage C++. Cependant, la délimitation d'un ensemble de fonctionnalités graphiques dans une interface utilisateur ne semble pas déclencher le désir d'explorer plus en profondeur les potentialités de l'outil. On constate aussi qu'un dialogue soutenu entre un praticien théâtral non-programmeur et un artiste numérique programmeur ne garantit pas une appropriation des ressorts fondamentaux de la représentation informatique du réel. Mon évolution personnelle en tant que chercheur et artiste m'a lentement conduit à reconsidérer mon rapport aux outils créatifs. Je peux ainsi mettre en regard mes expérimentations au démarrage de la mise en place d'AKN_Regie dans un dialogue avec Cédric Plessiet (Pluta, 2019) avec le choix de franchir le pas et d'essayer de comprendre directement le nouveau paradigme sur lequel les outils étaient fondés (Gagneré, 2020a). Une tension sous-jacente liée à la complexité des connaissances à mobiliser pour comprendre les différentes perspectives d'AKN_Regie est lisible dans les difficultés à s'approprier les résultats scéniques produits par l'outil. Gagneré et Plessiet (2019) ont décomposé les pièges qui peuvent surprendre une équipe de création pourtant complètement impliquée dans l'appropriation des enjeux numériques du travail avec des avatars.

Il semble ainsi que certaines résistances du théâtre à construire une relation constructive avec les possibilités expressives offertes par les nouvelles technologies et notamment l'informatique, repose sur des difficultés de représentation du réel, qui sont aussi présentes dans le domaine de la représentation des connaissances dans les écosystèmes informationnels. La construction d'un dispositif comme AvatarStaging et plus spécifiquement de l'outil AKN_Regie pour faciliter l'appropriation d'avatars comme entités scéniques offre un exemple de cette difficulté. Pouvoir déployer une allégorique englobant plusieurs perspectives pour s'approprier l'outil pourrait optimiser son utilisation créative en accord avec les potentialités offertes par l'informatique. La question qui demeure est donc celle de poursuivre le développement d'AKN_Regie en vue d'en faire une véritable passerelle entre des pratiques artistiques très différentes, et plus spécifiquement comment contribuer à un enrichissement de l'intelligence collective dans le développement culturel complexe de la société technologique contemporaine.

## 7 Références

Cycling74 (2023). https://cycling74.com/products/max (consulté le 7 février 2023).

Descola, P. (2005). *Par-delà nature et culture*. Gallimard.

Didascalie.net (2023). http://avatarstaging.eu (consulté le 7 février 2023).

Epic Games (2023a). https://www.unrealengine.com (consulté le 7 février 2023).

Epic Game (2023b). https://docs.unrealengine.com/5.1/en-US/blueprints-visual-scripting-in-unreal-engine/ (consulté le 7 février 2023).

Féral, J. (Eds.). (2018) *Corps en scène : Les acteurs face aux écrans*. L'Entretemps éditions (coll. « Les voies de l'acteur »).

Gagneré, G. (2015). Émergence et fragilité d'une recherche-création (2000-2007). In *Ligeia dossiers sur l'Art*, XXVIIIe année, n°137-140, 148-158, « Théâtres laboratoires : recherche-création et technologies dans le théâtre aujourd'hui », Losco Lena, M., Pluta, I. (Eds.).

Gagneré, G. (2020a). Du théâtre à l'informatique - Bascule dans un nouveau monde. In *JIT2020 - Journées d'Informatique Théâtrale*, février 2020, Performance Lab, Université Grenoble Alpes, Grenoble, hal-03419978.

Gagneré, G. (2020b). *The Shadow*. In *Proceedings of the 7th International Conference on Movement and Computing* (MOCO'20), Association for Computing Machinery, New York, NY, USA, 2020, Article 31, 1–2.

Gagneré, G. (2022). Le temps réel du temps réel. In *Scènes numériques. Anthologie critique. Digital Stages. Critical Anthology*, Pluta, I. (Eds.), Presse universitaires de Rennes.

Gagneré, G. (2023). AKN_Regie, un plugin dans Unreal Engine pour la direction d'avatar sur une scène mixte. In *JIT 2022 - Journées d'Informatique Théâtrale*, Oct. 2022.

Gagneré, G., Plessiet, C. (2015). Traversées des frontières. In *Frontières numériques & artéfacts*, Hachour H., Bouhaï N., Saleh I. (Eds.), L'Harmattan, Chapitre 1, 9-35.

Gagneré, G., Plessiet, C. (2016). Perceptions (théâtrales) de l'augmentation numérique. In *Actes du colloque international « Frontières Numériques : Perceptions »*, Toulon, Décembre, hal-02101604.

Gagneré, G., Plessiet, C. (2018). Experiencing avatar direction in low cost theatrical mixed reality setup. In *Proceedings of the 5th International Conference on Movement and Computing* (MOCO'18), Association for Computing Machinery, New York, NY, USA, Article 55, 1–6.

Gagneré, G., Plessiet, C. (2019). Espace virtuel interconnecté et Théâtre (2). Influences sur le jeu scénique. In *Revue : Internet des objets*, Numéro 1, Volume : 3, Février, ISSN : 2514-8273, ISTE OpenScience.

Gagneré, G., Plessiet, C. (2023). Quand le jeu vidéo est le catalyseur d'expérimentations théâtrales (2014-2019). In *Le jeu vidéo au carrefour de l'histoire, des arts et des médias*, Devès, C. (Eds.), Lyon, Les Éditions du CRHI, 209-219.

Gamma, E., Helm, R., Johnson, R., Vlissides, J. (1994). *Design Patterns. Elements of Reusable Object-Oriented Software,* Addison-Wesley.

Jacobson, I., Booch, G., Rumbaugh, J. (2000). *Le processus unifié de développement logiciel* (trad. de l'anglais par Zaim, V.), Paris : Eyrolles.

Lévy, P. (2010). De l'émergence de nouvelles technologies intellectuelles. In *Technologies de l'information et intelligences collectives*, Hermes Science Publications, 105.

Ong, W. J. (1982). *Orality and Literacy. The Technologizing of the Word*, Routledge.

Picon-Vallin, B. (Eds) (1998). *Les écrans sur la scène*, L'Âge d'Homme.

Pluta, I. (2019). When Theater Director Collaborates with Computer Engineer. In *Emerging Affinities - Possible Futures of Performative Arts*, Borowski, M., Chaberski, M., Sugiera, M. (Eds.), Transcript Verlag, Bielefeld, 49-70.

Simondon, G. (1958 édition originale). *Du mode d'existence des objets techniques*, Éditions Jérôme Millon, Grenoble, 2005, 309-310.

Szoniecky, S. (2012). *Évaluation et conception d'un langage symbolique pour l'intelligence collective : Vers un langage allégorique pour le Web*, Sciences de l'information et de la communication, Thèse de l'Université Paris 8, PARAGRAPHE https://theses.hal.science/tel-00764457.

Ternova, A., Gagneré, G. (2023). Le potentiel créatif du plugin AKN_Regie dans un contexte théâtral. In *JIT 2022 - Journées d'Informatique Théâtrale*, Oct. 2022, Lyon.

Triffaux, J.-P. (2008). Le comédien à l'ère numérique. In *Communications* 2008/2 (n° 83), 207.

Valéro, J., Ronfard, R. (2020). Pourquoi l'informatique théâtrale ?. In *JIT 2020 - Journées d'Informatique Théâtrale*, Performance Lab, Université Grenoble Alpes, Février 2020, Grenoble.